\documentclass[twocolumn]{openjournal}

\usepackage{xcolor}
\usepackage{textgreek}
\usepackage[utf8x]{inputenc}
\usepackage[english]{babel}
\usepackage{amsmath}
\let\oldAA\AA
\renewcommand{\AA}{\text{\normalfont\oldAA}}

\usepackage{hyperref}
\hypersetup{
    colorlinks=true,
    linkcolor=blue,
    citecolor=blue,
    filecolor=blue,
    urlcolor=blue,
}
\usepackage{tensind}
\tensordelimiter{?}
\DeclareGraphicsExtensions{.bmp,.png,.jpg,.pdf}
\usepackage{verbatim}
\usepackage[normalem]{ulem}
\usepackage{orcidlink}
\usepackage{soul}
\usepackage{appendix}
\graphicspath{ {./figs/} }

\begin{document}

\title{Morphological fingerprints of Forbush Decreases and their relation to geomagnetic storm severity}

\author{Juan D. Perez-Navarro$^{a,*}$ \orcidlink{0009-0008-5689-314X}}
\email[Juan D. Perez-Navarro: ]{perezjuan@utb.edu.co}
\affiliation{Universidad Tecnológica de Bolívar. Escuela de Transformación Digital. Parque Industrial y Tecnológico Carlos Vélez Pombo Km 1 Vía Turbaco. Cartagena de Indias, 130010, Colombia}
\affiliation{$^{*}$ Corresponding author: perezjuan@utb.edu.co (Juan D. Perez-Navarro)}

\author{D. Sierra-Porta$^{a,**}$ \orcidlink{0000-0003-3461-1347}}
\email[D. Sierra-Porta: ]{dporta@utb.edu.co}
\affiliation{Universidad Tecnológica de Bolívar. Grupo de Investigación Gravitación y Matemática Aplicada - GIGMA \& Grupo de Investigación Física Aplicada y Procesamiento de Imágenes y Señales - FAPIS. Parque Industrial y Tecnológico Carlos Vélez Pombo Km 1 Vía Turbaco. Cartagena de Indias, 130010, Colombia}
\affiliation{$^{**}$ Corresponding author: dporta@utb.edu.co (D. Sierra-Porta)}

\begin{abstract}
Forbush decreases (FDs) are transient depressions in the galactic cosmic-ray flux observed by global neutron-monitor networks and are commonly associated with interplanetary disturbances driven by coronal mass ejections and related shocks. Despite extensive observational work, quantitatively comparing FD morphology across events and linking it to storm severity remains challenging due to heterogeneous station responses, coverage gaps, and the multivariate nature of the network. This work introduces a graph-based event representation in which each FD is mapped to an event network constructed from pairwise dissimilarities between station response time series. A controlled sparse backbone is obtained via the minimum spanning tree, enabling comparable event graphs across cases. From each graph, a compact set of geometric/topological fingerprints is computed, including global integration measures, spectral summaries, mesoscopic structure, centrality aggregates, and complexity descriptors. 

Predictive skill is assessed using strict leave-one-event-out validation over a pre-defined grid of distance metrics and distance-domain transformations, with selection criteria fixed \emph{a priori}. The proposed fingerprints exhibit measurable signal for three tasks: (i) multi-class classification of geomagnetic storm intensity (G3/G4/G5) with moderate but consistent performance and errors dominated by adjacent categories; (ii) stronger binary severity screening ($\ge$G4 vs.\ G3) with high sensitivity to severe events; and (iii) drop regression with partial least squares achieving positive explained variance relative to a fold-wise mean baseline. These results support the premise that FD event morphology leaves reproducible network signatures that can be captured by compact graph fingerprints, offering a complementary, interpretable framework for comparative FD characterization and for studying how heliospheric drivers imprint structure on global neutron-monitor responses.
\end{abstract}


%

\section{Introduction}
\label{sec:intro}

Forbush decreases (FDs) are transient reductions in the galactic cosmic-ray (GCR) flux observed at Earth \citep{forbush1937effects, 2022arXiv221213514C}. The physical mechanism behind Forbush decreases involves the magnetic field of plasma solar wind sweeping galactic cosmic rays away from Earth \citep{2007astroph1860G, 2000SSRv9355C}. These events are associated with large-scale perturbations in the solar wind and interplanetary magnetic field, particularly non-recurrent decreases that are {caused by transient interplanetary events related to coronal mass ejections (CMEs) from the Sun and their interplanetary counterparts (ICMEs)} \citep{2013SoPh284167B, 2017arXiv171000945H}. 

{It is worth emphasizing that the physical relationship between FD amplitude and geomagnetic storm severity is indirect: while both phenomena are driven by CME/ICME transients, storm intensity is particularly sensitive to the southward component of the interplanetary magnetic field ($B_z<0$), whereas FD amplitude depends primarily on the magnetic barrier structure of the ICME sheath and flux rope, which modulates cosmic-ray access independently of $B_z$ polarity \citep{NOAA_SWPC_Gscale, 2000SSRv9355C}.}

Because neutron monitors register the integrated response of secondary particles generated in the atmosphere, FD signatures are routinely tracked in multi-station networks, and they remain a standard observable for studying the coupling between interplanetary transients and near-Earth particle environments \citep{2025arXiv250617917C, 2011JPhCS287a2034B, 2013JPhCS409a2202P}. Despite decades of observational work, robust event-level characterization across stations continues to face practical challenges: FD morphologies vary with event type, viewing geometry, background conditions, and data completeness, while station responses differ systematically across the global network. {Recent work has further characterized FDs using global neutron monitor networks \citep{2025arXiv250617917C}, muon telescope arrays \citep{2011JPhCS287a2034B}, anisotropy measurements and leader fraction analysis \citep{2013JPhCS409a2202P, 2008JASTP70207P, 2016ApJ81738R}, and precursor signatures such as loss-cone enhancements and pre-event flux variations.}

A key driver of inter-station heterogeneity is geomagnetic shielding. Station cutoff rigidity determines the minimum particle rigidity required to reach a given detector and therefore modulates both FD amplitude and timing \citep{2026AdSpR773549Y}. As a result, FD strength is not solely an intrinsic property of the interplanetary driver; it is also filtered by the spatial distribution of observing stations and their rigidity thresholds \citep{2021ApJ9069L, 2019ICRC361084I, 2025afasconfE27O, 2022PhyA60728159S, 2024Chaos34b3114S}. {Beyond cutoff rigidity, the asymptotic viewing direction of each station — which determines the portion of the sky sampled by the detector — introduces additional inter-station variability in FD onset timing and amplitude, {since stations whose asymptotic directions become magnetically connected to the disturbed interplanetary region may register a precursor signature earlier than others \citep{smart2005review}.}

This motivates analyses that explicitly combine multi-station measurements with station metadata rather than treating each time series in isolation. In practice, however, many studies focus on a limited number of stations, compress the network into summary curves, or rely on pairwise comparisons without a unifying representation that is directly comparable across events.

Graph-based representations provide a principled way to summarize multivariate observations \citep{2023arXiv230102333F, 2021arXiv211009887F}. Nodes can represent stations and edges can encode inter-station similarity or dissimilarity, yielding an event-specific network that can be characterized by well-defined structural descriptors. In the context of FD observations, such representations offer a pathway to convert multi-station time series into compact event fingerprints that can be compared across drivers, storm intensities, and event magnitudes \citep{2025arXiv251101506P}. A central methodological issue is comparability: naive similarity thresholds often generate overly dense graphs that trivialize many network metrics, whereas overly aggressive pruning can distort the underlying organization \citep{2017arXiv170510817G}. A reproducible graph-based FD framework must therefore specify (i) the event window and missing-data treatment, (ii) the similarity/distance definition, and (iii) a graph-construction rule that yields interpretable and comparable connectivity across events.

This work introduces an event-level network construction for FDs observed in the Neutron Monitor Database (NMDB). For each event, stations are treated as nodes and edges are derived from pairwise dissimilarities between station response series within a common event window, with missing values handled through alternative, explicitly defined strategies (coverage-based station filtering or iterative imputation). In addition, transformations and optional scalings are applied in the distance domain to control dynamic range and enhance cross-event comparability. For visualization and controlled backbone comparison, minimum spanning trees (MST) are used to provide sparse \citep{stam2014trees}, connected representations that highlight the dominant inter-station couplings while keeping the edge count fixed across events.

The scientific objective is to test whether graph fingerprints capture systematic differences between FD events stratified by geomagnetic storm intensity and by FD magnitude (drop). Three complementary predictive tasks are considered: multi-class intensity classification (G3/G4/G5), binary severity screening ($\ge$G4 vs.\ G3), and drop regression. {Throughout this work, storm intensity is classified using the NOAA geomagnetic storm scale (G-scale), where G1 denotes minor storms and G5 denotes extreme events \citep{NOAA_SWPC_Gscale}.} Evaluation is performed under strict leave-one-event-out validation, and model selection is restricted to a pre-defined, theory-motivated grid of pipeline choices with selection criteria fixed \emph{a priori}.

The contributions of this study are threefold. First, it provides a reproducible pipeline that maps multi-station FD time series into event graphs with explicit control of preprocessing, missing-data handling, and distance-domain operations. Second, it defines a compact set of graph descriptors spanning integration, spectral structure, mesoscopic organization, centrality aggregates, and complexity measures, enabling event-level analysis in small-sample regimes with leakage-free validation. Third, it provides evidence that graph fingerprints derived from NMDB carry information about storm severity and FD magnitude and reveal rigidity-conditioned organization in representative event backbones, supporting graph-based characterization as a complementary tool for FD and space-weather studies.

\section{Data}
\label{sec:data}
{We use hourly neutron-monitor count-rate records} from the NMDB, selecting the subset of stations available during each Forbush decrease event \citep{2023dashconfE19S, 2016AGUFMIN44A05S, 2011AdSpR472210M}. For each event, we extract a common analysis window around the disturbance and assemble a time-aligned matrix whose columns correspond to stations and whose rows correspond to measurement times. Station coverage varies across events due to data availability and quality control.

{To characterize the magnitude of the event at each station, we compute the percentage reduction in the cosmic-ray count rate with respect to a pre-event reference level defined as the mean count rate over a quiet period of 24--48 hours immediately preceding the FD onset. This reference is initially derived from NMDB event metadata and subsequently verified by visual inspection of each individual time series to confirm the absence of prior disturbances.}

{Each FD event is assigned a storm intensity class on the NOAA G-scale, compiled from widely used FD catalogs and peer-reviewed studies, complemented with operational space-weather reports from NASA and GOES-based products where applicable. The complete list of events, including dates, storm class, observed FD drop, and data sources, is provided in Appendix~\ref{app:events}. A full description of the neutron monitor stations used — including cutoff rigidity, geographic coordinates, altitude, and mean count rate — is given in Appendix~\ref{app:stations}.}

Finally, we collect station metadata from NMDB and associated station documentation, including cutoff rigidity, geographic coordinates (latitude and longitude), and altitude. These attributes are used both for descriptive analysis and to evaluate rigidity-conditioned network structure and node-level roles across events. {In addition to cutoff rigidity, we note that the asymptotic viewing direction of each station introduces further variability in FD onset timing across the network; this geometric effect is discussed in the context of the results in Section~\ref{sec:discussion}.}

\section{Methods}
\label{sec:methods}

\subsection{Event windows and missing-data treatments}
\label{sec:preproc}
For each Forbush decrease (FD) event, a common analysis window is extracted from multi-station NMDB count-rate records and arranged into a time-aligned matrix $X \in \mathbb{R}^{T \times N}$, where $T$ is the number of time samples and $N$ is the number of stations available for that event. 

{Count-rate records are used at hourly resolution, consistent with the temporal scale of Forbush decrease dynamics. Where stations provide only minute-resolution data, hourly rates are obtained by averaging, which both reduces statistical noise and allows the inclusion of additional stations that report only at hourly cadence.} A complementary missing-data treatments is considered to assess robustness. (i) Coverage filtering: stations whose fraction of missing values within the event window exceeds a fixed threshold $\tau=0.5$ are excluded, and subsequent computations operate on the retained station set. (ii) Imputation: {missing samples in $X$ are imputed using an iterative multivariate scheme with a $k$-nearest neighbours ($k$-NN, $k=2$) regressor as the base estimator \citep{van2011mice, pedregosa2011scikit}, producing a complete matrix prior to distance estimation.} These two treatments define alternative event representations and are treated as part of the pre-defined pipeline grid.

\subsection{Pairwise distances between station responses}
\label{sec:similarity}
Let $\mathbf{x}_i \in \mathbb{R}^{T}$ denote the response series for station $i$ in a given event window. {Each series $\mathbf{x}_i$ {represents the percentage deviation of the hourly count rate} from the station-specific pre-event reference level, as defined in Section \ref{sec:data}. This representation places all stations on a common relative scale, making pairwise distances directly comparable across stations with very different absolute count rates (which depend on detector size, altitude, and tube configuration).} Event-specific pairwise distances are computed between station responses to form a matrix $D \in \mathbb{R}^{N \times N}$ with entries $d_{ij}$. Two $\ell_p$ distances are considered:
\begin{equation}
d_{ij}^{(p)} = \left(\sum_{t=1}^{T} \left|x_i(t)-x_j(t)\right|^p \right)^{1/p},
\end{equation}
where $p=1$ corresponds to Manhattan distance, and $p=2$ corresponds to the Euclidean member of the Minkowski family \citep{2015PLoSO1044059S, 2019arXiv190502257W, 2021arXiv210210231S}. This choice provides a parsimonious comparison between a more robust absolute-deviation geometry ($p=1$) and a quadratic geometry ($p=2$).

{We note that this percentage-deviation representation does not fully equalize station responses: stations at higher cutoff rigidity exhibit systematically smaller percentage variations (both from FDs and from other modulation, with the notable exception of the diurnal variation, which is nearly rigidity-independent). An additional standardization step --- e.g., per-station $z$-scoring of $\mathbf{x}_i$ prior to distance computation --- could place all stations on a more equal footing and is a natural candidate for the pipeline grid in future work.}

\subsection{Distance-domain transformations and normalizations}
\label{sec:dist_transform}
To control dynamic range and improve comparability across events, additional operations are applied directly in the distance domain. Let $t$ denote a robust scale defined as the median of the off-diagonal distances $\{d_{ij}\}_{i<j}$ for a given event. Three distance-domain transformations are explored:
\begin{align}
\tilde d_{ij} &= d_{ij} \quad \text{(none)}, \\
\tilde d_{ij} &= \log\!\left(\frac{d_{ij}}{t}\right) \quad \text{(log)}, \\
\tilde d_{ij} &= \exp\!\left(\frac{d_{ij}}{t}\right) \quad \text{(exponential)}.
\end{align}
After transformation, an optional normalization is applied to $\tilde D$ (e.g., none, min-max, z-score, robust scaling, or decimal scaling), yielding a final event-specific dissimilarity matrix used for graph construction. 

{Each transformation serves a distinct purpose. The identity (no transform) preserves absolute dissimilarity magnitudes, which may retain information about event amplitude. The log transform compresses the dynamic range of the distance matrix, reducing the influence of outlier station pairs and stabilizing topology- and centrality-based summaries across events with heterogeneous station coverage. The exponential transform amplifies large dissimilarities, sharpening contrasts between strongly and weakly coupled station pairs at the cost of increased sensitivity to outliers.}

{The optional post-transformation normalization can introduce systematic effects that interact with downstream graph metrics. Min-max scaling maps all events to the same distance range $[0,1]$, removing cross-event amplitude information but improving comparability of graph topology. Z-score and robust scaling preserve relative structure while standardizing spread; decimal scaling reduces numerical magnitude without altering ratios. These effects are part of the explicitly modeled pipeline grid and are accounted for through the \emph{a priori} model selection protocol described in Section~\ref{sec:evaluation}.}

\subsection{Graph construction}
\label{sec:graph}
Each event is represented as a weighted, undirected graph $G=(V,E,W)$ \citep{chung2006complex, kim2008complex} where nodes correspond to stations ($|V|=N$). Edges connect all station pairs ($E$ is complete over the retained stations), and weights are given by the transformed/normalized dissimilarities ($w_{ij}=\tilde d_{ij}$). In this representation, smaller weights indicate stronger similarity (shorter dissimilarity). No additional edge sparsification is imposed; comparability across events is primarily enforced through the station-level coverage filtering and the distance-domain scaling described above.

A consistent visual pattern is a core--periphery organization in which intermediate-rigidity stations tend to occupy bridging positions, while very low-rigidity stations (polar/high-latitude, darker colors) more often appear as peripheral branches. High-rigidity stations (warmer colors) show partial clustering and connect through a small number of intermediates, suggesting that similarity in rigidity can translate into similarity of FD response morphology and, consequently, shorter dissimilarities. This qualitative behavior motivates rigidity-aware summaries such as assortativity and centrality--rigidity relations at the event level.

Figure~\ref{fig:example_graph} provides an illustrative event-level network visualization for a representative FD. Nodes correspond to neutron-monitor stations and node color encodes the cutoff rigidity (GV). Edges correspond to the MST computed from the event-specific dissimilarity matrix, providing a sparse, connected backbone with fixed edge count ($N-1$) and enabling direct visual comparison across distance metrics.

The two vertical panels highlight that changing the distance metric (Manhattan vs.\ Minkowski) can alter the backbone geometry (e.g., hubness and branch structure) while preserving the main large-scale station groupings.

\begin{figure*}[htb]
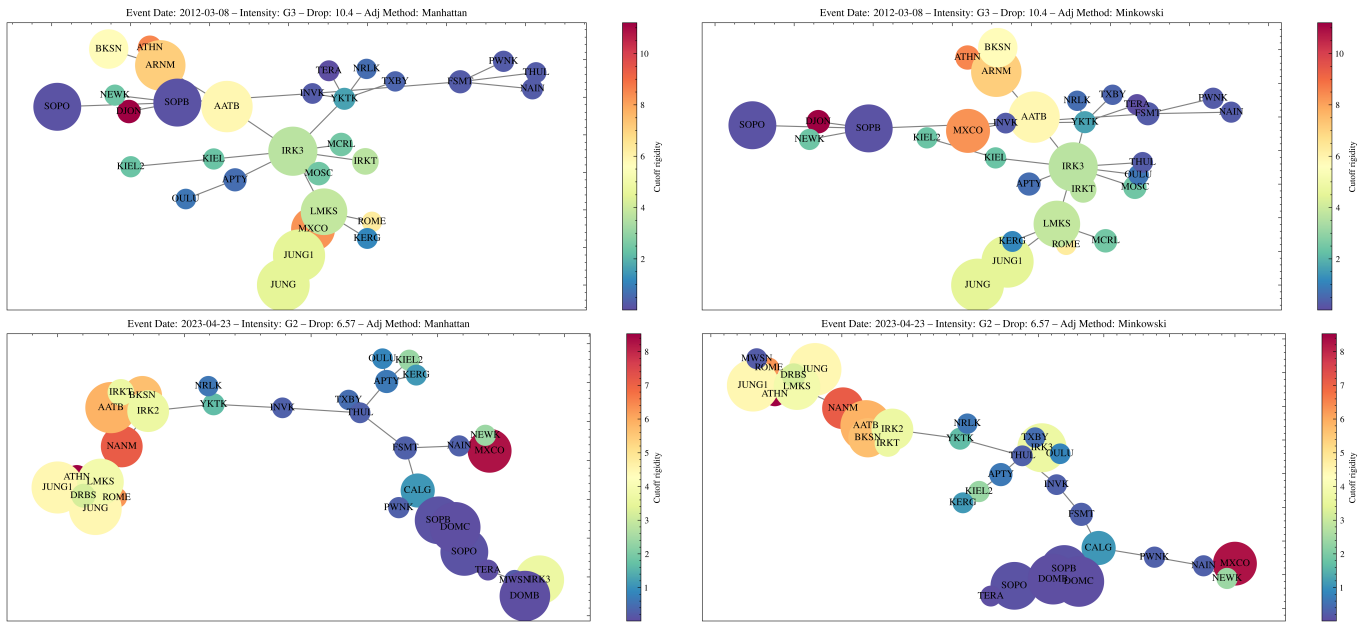

\centering
\includegraphics[width=\textwidth]{20120308_graph-altitude.png}
\includegraphics[width=\textwidth]{20230423_graph-altitude.png}
\caption{Illustrative two event graph for 2012-03-08 (G3) and 2023-04-23 (G2) {which are representative Forbush decreases}. Nodes are stations colored by cutoff rigidity (GV). Edges correspond to the MST computed from the event-specific dissimilarity matrix, yielding a sparse, connected backbone with $N-1$ edges that minimizes total distance. Left: Manhattan ($p=1$); right: Minkowski/Euclidean ($p=2$). While the main station groupings persist, the MST topology (branching patterns and connector nodes) changes with the distance metric.}
\label{fig:example_graph}
\end{figure*}

For the illustrated event (drop $\approx 10.4\%$), the backbone already exhibits coherent connectivity across multiple rigidity regimes, but with visible heterogeneity in node roles: some stations act as connectors across subgroups, while others remain weakly coupled and attach as leaves. Across the full event set, the regression analysis formally tests whether such global organization patterns (e.g., efficiency, spectral complexity, centrality aggregation, and mixing structure) carry predictive signal for drop magnitude beyond a fold-wise baseline, rather than relying on individual visual examples.

{For visualization and controlled backbone comparison, the minimum spanning tree (MST) is computed from the event-specific dissimilarity matrix using Kruskal's algorithm \citep{kruskal1956shortest}. The MST retains exactly $N-1$ edges --- the minimal connected subgraph that links all $N$ stations while minimizing total edge weight. This produces a sparse, connected backbone that highlights the dominant inter-station dissimilarity structure while keeping the edge count fixed and independent of event-specific threshold choices. Cross-event comparability is thereby enforced by construction: all event graphs share the same topological constraint ($N-1$ edges), and differences in fingerprints reflect genuine differences in inter-station organization rather than artifacts of density variation \citep{stam2014trees}.}

\subsection{Event-level graph fingerprints}
\label{sec:features}
From each weighted event graph, a compact fingerprint vector is computed by aggregating node-level and spectral/mesoscopic properties into event-level descriptors. The reported feature set includes: global efficiency, Estrada index, modularity, assortativity, average Katz centrality, average closeness, average betweenness, and a Laplacian-based summary, together with {additional complexity descriptors: a Shannon entropy of the edge-weight distribution, a fractal-dimension estimate obtained via the Compact-Box-Burning (CBB) algorithm \citep{song2007calculate} applied directly to the MST graph, varying the box-diameter threshold $\ell_b \in \{2,\ldots,11\}$, counting the minimum number of boxes required to cover all nodes at each scale, and extracting the slope of the $\log N_b$ vs.$\log \ell_b$ regression (so that $d_f$=-slope); and a Hurst exponent estimated via rescaled-range (R/S) analysis \citep{nolds} applied to the ordered sequence of MST edge weights.}

When centrality measures depend on path lengths, weighted shortest paths are computed using the dissimilarity weights. Event-level centralities are obtained by averaging node-wise values, producing comparable scalars across events. These descriptors are designed to capture complementary aspects of global integration (efficiency), spectral complexity (Estrada), mesoscopic structure (modularity), mixing patterns (assortativity), and heterogeneity/complexity (entropy/fractal/Hurst) in the station-response network. For a description of these descriptors, metrics and markers see for example \cite{omar2020survey} and references therein.

\subsection{Event labels, predictive tasks, and evaluation protocol}
\label{sec:evaluation}
Each FD event is assigned (i) a storm intensity label on the NOAA G-scale and (ii) a continuous magnitude label given by the observed FD drop (\%). Three tasks are considered: multi-class intensity classification (G3/G4/G5), binary severity screening ($\ge$G4 vs.\ G3), and drop regression.

Model selection is performed over a pre-defined, theory-motivated grid of pipeline choices (missing-data treatment, distance metric, distance-domain transformation, and optional normalization). Selection criteria are fixed \emph{a priori}: macro-F1 for classification and $R^2$ for regression. Predictive evaluation is carried out with strict leave-one-event-out (LOEO) validation. For regression, performance is additionally compared against a fold-wise mean-predictor baseline computed on each training fold.

{In LOEO, each of the $n$ events is held out in turn as the test set while the remaining $n-1$ events form the training set; this process is repeated $n$ times and performance is aggregated across all held-out events. Any preprocessing steps that involve data-driven fitting --- specifically, median imputation and feature standardization --- are fitted exclusively on the training fold and applied to the held-out event, ensuring that no information from the test event leaks into the model. Given the modest sample size ($n\leq 34$), LOEO provides the least biased estimate of generalization error available without external validation data.}

{This evaluation design follows established practice for small-sample classification in space-physics contexts \citep{2024Chaos34b3114S, 2022PhyA60728159S}.}

\section{Results}
\label{sec:results}
{The model-selection grid spans distance-domain transformations, normalization choices, and two adjacency metrics, evaluated under strict LOEO with leakage-free preprocessing. Selection criteria were fixed \emph{a priori} (macro-F1 for classification; $R^2$ for regression). Only the best-performing configuration per task is reported here; the full sweep is available in the accompanying repository.}

\subsection{Rigidity-conditioned node roles in event graphs}
\label{sec:results_rigidity_roles}

To assess whether geomagnetic shielding shapes \emph{local} station roles within event graphs, node-level descriptors were stratified by cutoff rigidity and summarized across events. Stations were grouped into three rigidity bands using fixed physical thresholds: low rigidity ($R_c < 3$~GV), high rigidity ($R_c \ge 6$~GV), and an intermediate group spanning $3 < R_c < 6$~GV. For each event, node metrics were aggregated within each band to obtain group-level summaries, and the resulting distributions across events were compared.

Figure~\ref{fig:rigidity_boxplots} shows the across-event distributions for four representative descriptors (avg. katz, avg. closeness, avg. betweenness, and a Laplacian-based summary). A clear separation by rigidity emerges. Low-rigidity stations exhibit systematically larger avg. katz and avg\_closeness across events, indicating that these stations tend to occupy more central positions in the event-network backbone and contribute disproportionately to global communicability-like structure. In the same group, avg. betweenness is also markedly higher, suggesting that low-rigidity stations more frequently act as bridges connecting subnetworks (or branches) in the backbone representation. Conversely, medium- and high-rigidity groups show substantially lower centrality levels, consistent with a more peripheral or locally confined role.

{The Laplacian-based summary displays a complementary pattern: the intermediate-rigidity group ($3 \le R_c < 6$~GV) exhibits larger values and broader variability than the low- and high-rigidity groups. This is physically interpretable in terms of the shielding structure of Earth's magnetosphere: low-rigidity (polar) stations are most sensitive to GCR modulation and respond strongly to any interplanetary barrier, making them dominant hubs in the dissimilarity backbone; intermediate-rigidity stations, partially shielded, mediate between the strongly modulated polar responses and the weakly modulated equatorial responses, and consequently tend to occupy spectral-bridging positions in the graph; high-rigidity stations are the least modulated and appear as peripheral nodes. This hierarchy --- directly traceable to geomagnetic shielding --- is precisely what makes rigidity-conditioned centrality a physically grounded fingerprint component. Notably, the two descriptors showing the strongest rigidity separation here (avg. Katz and avg. Laplacian) also rank among the top predictors in the intensity classification models (Table~\ref{tab:feat_importance}), directly linking physical interpretability to predictive performance.}

\begin{figure}[t]
\centering
\includegraphics[width=\columnwidth]{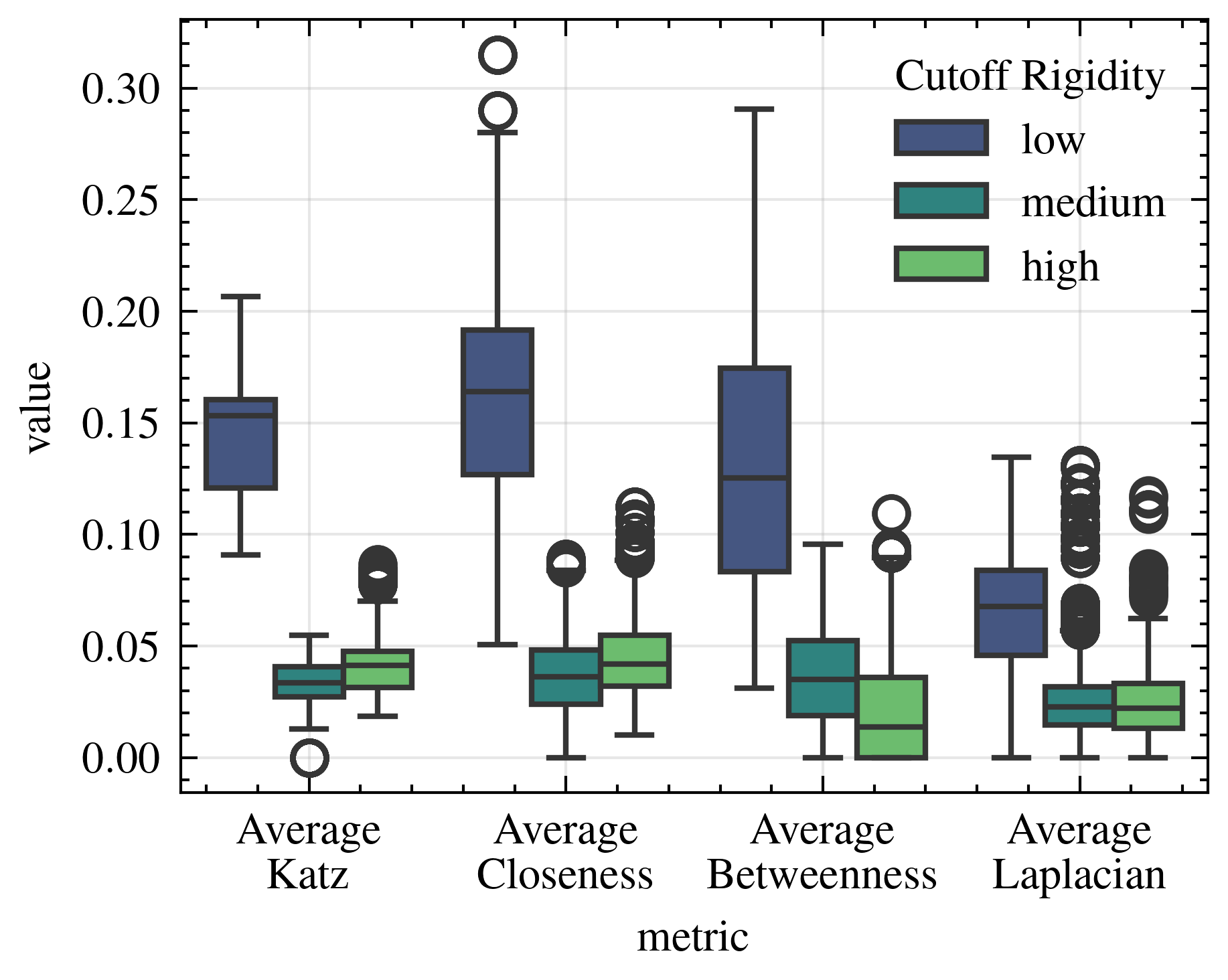}
\caption{Rigidity-conditioned distributions of node-role summaries across events. For each FD event graph, node-level metrics are aggregated within three cutoff-rigidity bands defined by fixed thresholds: low ($R_c < 3$~GV), medium ($3  \le R_c < 6$~GV), and high ($R_c \ge 6$~GV). Boxplots summarize the distributions across events. Low-rigidity stations show systematically larger avg. katz, avg. closeness, and avg. betweenness, indicating more central and bridging roles in event backbones, while the Laplacian-based summary peaks for the medium-rigidity group.}
\label{fig:rigidity_boxplots}
\end{figure}

\subsection{Multi-class classification of storm intensity (G3/G4/G5)}
\label{sec:results_multiclass}

The main classification task targets the categorical storm intensity associated with each Forbush event (G3/G4/G5). The best multi-class result was obtained using a log transformation in the distance domain, no post-distance normalization, and Minkowski-based adjacency, followed by linear discriminant analysis. Under LOEO ($n=33$ events), the classifier achieved an accuracy of $0.576$, balanced accuracy of $0.572$, and macro-F1 of $0.575$.

Table~\ref{tab:cm_multiclass} reports the confusion matrix. Class-wise recalls are $\mathrm{Rec}_{G3}=6/10=0.60$, $\mathrm{Rec}_{G4}=8/13=0.62$, and $\mathrm{Rec}_{G5}=5/10=0.50$, indicating that performance is not driven by a single class. Misclassifications are dominated by neighboring categories (notably G4$\leftrightarrow$G5), consistent with the expectation that storm severity forms a continuum and that graph fingerprints may partially overlap when intensities are close in magnitude. Figure~\ref{fig:lda_proj} visualizes the same configuration in discriminant space (LD1--LD2), showing a moderate but nontrivial organization by intensity.

\begin{table}[htb]
\centering
\caption{LOEO confusion matrix for the best multi-class intensity model (LDA; G3/G4/G5; $n=33$).}
\label{tab:cm_multiclass}
\begin{tabular}{lccc}
\hline
 & \textbf{Pred G3} & \textbf{Pred G4} & \textbf{Pred G5} \\
\hline
\textbf{True G3} & 6 & 1 & 3 \\ \hline
\textbf{True G4} & 1 & 8 & 4 \\ \hline
\textbf{True G5} & 2 & 3 & 5 \\
\hline
\end{tabular}
\end{table}

\begin{figure}[htb]
\centering
\includegraphics[width=\columnwidth]{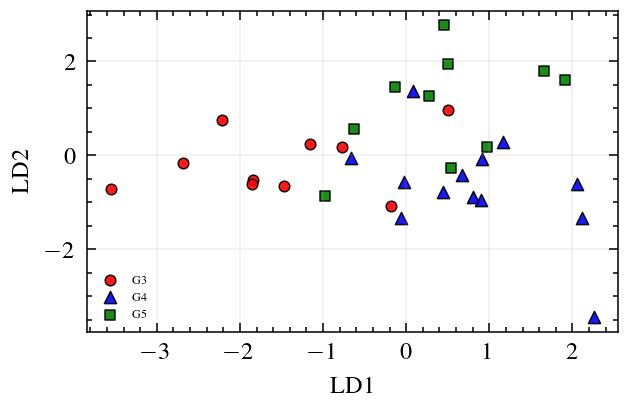}
\caption{LDA projection (LD1--LD2) for the best multi-class intensity pipeline (log distance transform, no normalization and Minkowski adjacency). Colors indicate storm classes (G3/G4/G5).}
\label{fig:lda_proj}
\end{figure}

\subsection{Binary severity screening: $\ge$G4 versus G3}
\label{sec:results_binary}

{A binary severity screening task was also considered, grouping strong storms (G4/G5) versus moderate storms (G3). The best-performing configuration for this task --- log distance transform, decimal-scaling normalization, and Minkowski adjacency --- differs from the multiclass optimum, reflecting that the binary and multiclass tasks favor different representations of inter-station dissimilarity structure. Under LOEO ($n=33$), LDA achieved an accuracy of 0.788, balanced accuracy of 0.735, and macro-F1 of 0.741 (Table~\ref{tab:best_summary}).}

Sensitivity to severe events is high: the true positive rate for $\ge$G4 is $20/23=0.87$. Specificity for G3 is lower ($5/10=0.50$), indicating that a subset of moderate events remains difficult to separate from the severe group. This asymmetry is consistent with the expectation that stronger disturbances imprint more distinctive global network signatures, while moderate events can occupy an intermediate regime in which fingerprints overlap with adjacent categories.

\begin{table}[htb]
\centering
\caption{LOEO confusion matrix for binary severity screening (LDA; $\ge$G4 vs.\ G3; $n=33$).}
\label{tab:cm_binary}
\begin{tabular}{lcc}
\hline
 & \textbf{Pred G3} & \textbf{Pred $\ge$G4} \\
\hline
\textbf{True G3} & 5 & 5 \\ \hline
\textbf{True $\ge$G4} & 3 & 20 \\
\hline
\end{tabular}
\end{table}

\subsection{Drop regression with PLS (5 components)}
\label{sec:results_regression}

The regression task evaluates whether graph fingerprints encode quantitative information about Forbush magnitude by predicting the observed drop (\%). The best performance was obtained with Minkowski-based adjacency and no distance transformation or normalization, combined with partial least squares regression (PLS) using 5 components. Under LOEO ($n=34$), the model achieved $R^2=0.350$, MAE $=2.41$ percentage points, and RMSE $=3.06$.

This outperformed a fold-wise mean-predictor baseline (LOEO), which yielded $R^2=-0.062$, MAE $=2.69$, and RMSE $=3.91$. Figure~\ref{fig:pls_drop} shows predicted versus observed drops, indicating measurable skill together with a typical regression-to-the-mean pattern: very large drops are underpredicted and small drops slightly overpredicted, which is expected given the limited sample size and the intrinsic variability linking event morphology to magnitude.

\begin{figure}[htb]
\centering
\includegraphics[width=\columnwidth]{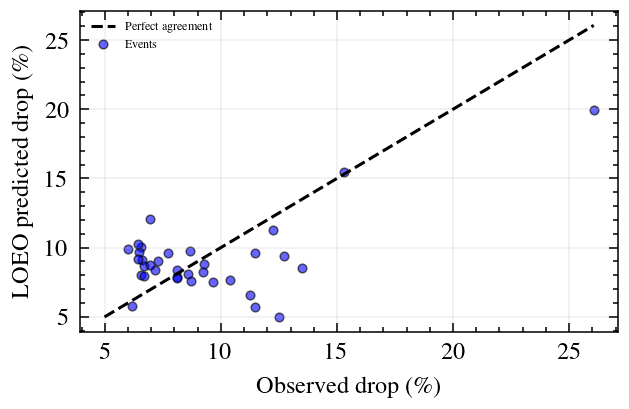}
\caption{{PLS (5 components) LOEO regression: predicted vs.\ observed Forbush drop (\%) for the best pipeline (no transform; no normalization; Minkowski adjacency). Each point represents one held-out event. The dashed line indicates perfect agreement between predicted and observed values.}}
\label{fig:pls_drop}
\end{figure}

\subsection{Summary of best configurations}
\label{sec:results_summary}

Table~\ref{tab:best_summary} summarizes the best-performing configurations and LOEO performance. Overall, the results support the hypothesis that geometric/topological graph fingerprints extracted from multi-station Forbush observations contain reproducible signal for (i) stratifying storm intensity and (ii) partially explaining drop magnitude. Given the modest number of events, these findings are best interpreted as evidence of signal in the proposed fingerprints rather than as a final operational model; larger event sets will enable tighter uncertainty quantification and more definitive benchmarking.

\begin{table}[htb]
\centering
\scriptsize
\setlength{\tabcolsep}{3.5pt}
\renewcommand{\arraystretch}{1.15}
\caption{Best-performing configurations (selected by LOEO performance) for each task.}
\label{tab:best_summary}
\begin{tabular}{p{0.23\columnwidth} p{0.34\columnwidth} p{0.34\columnwidth}}
\hline
\textbf{Task} & \textbf{Best setting} & \textbf{LOEO performance} \\
\hline
Multi-class intensity (G3/G4/G5) &
LDA; log distance transform; no normalization; Minkowski adjacency &
Acc = 0.576; Balanced Acc = 0.572; Macro F1 = 0.575 ($n=33$) \\
\hline
Binary severity ($\ge$G4 vs G3) &
LDA; log distance transform; Decimal-Scaling; Minkowski adjacency  &
Acc = 0.7878; Balanced Acc = 0.7347; MacroF1 = 0.7413 ($n=33$) \\
\hline
Drop regression &
PLS-5; no transform; no normalization; Minkowski adjacency &
$R^2$ = 0.350; MAE = 2.41; RMSE = 3.06 ($n=34$) \\
\hline
\end{tabular}
\end{table}

\subsection{Comparison with scalar baselines}
\label{sec:results_baselines}

{To assess the practical advantage of the graph-based pipeline over simpler
alternatives, we evaluated six scalar baselines under the same LOEO protocol and LDA classifier used for the graph fingerprints. Baselines include: the maximum FD drop observed across polar stations ($R_c < 3$~GV), {the minimum southward IMF $B_z$ component, the mean interplanetary magnetic field magnitude $|\mathbf{B}|$, and the mean solar-wind speed, each extracted from the OMNI hourly dataset over a window spanning 6~hours before to 24~hours after the event onset --- a duration consistent with the typical main-phase and early-recovery timescale of a geomagnetic storm; a two-dimensional feature vector formed by the joint pair ($B_z$ minimum, $|\mathbf{B}|$ mean) over the same window, used directly as input to the LDA classifier; and the minimum Dst index.} OMNI data were obtained from the GSFC/SPDF OMNIWeb interface \citep{king2005solar}.}

{Table~\ref{tab:baselines} summarizes the results. All scalar baselines fall
below the graph fingerprints across the three tasks. In the multiclass setting, {the best scalar/vector baseline is the joint ($B_z$, $|\mathbf{B}|$) pair (macro-F1 = 0.561), which lies 0.014 points below the graph fingerprints (macro-F1 = 0.575); the minimum southward $B_z$ alone reaches macro-F1 = 0.501. In the binary screening task, the same joint baseline reaches macro-F1 = 0.618 against 0.741 for the graph pipeline (+0.123).}

The polar max-drop — the most direct simple measure of FD amplitude — performs below the stratified-random baseline in the multiclass task (macro-F1 = 0.285 $<$ 0.333), confirming that raw amplitude at polar stations does not stratify storm intensity. For drop regression, all scalar baselines yield $R^2 < 0$, indicating they perform worse than a fold-wise mean predictor, whereas the graph PLS achieves $R^2 = 0.350$. {In absolute terms, the graph PLS achieves the only positive $R^2$ among all methods; while its MAE (2.41) is comparable to or better than most baselines, the polar max-drop baseline shows a similarly low MAE (2.28) despite its negative $R^2$ (-0.007), illustrating that low absolute error alone does not imply predictive skill above a trivial mean-predictor.}

{These results demonstrate that the geometric and topological structure encoded by the graph fingerprints carries information about storm severity and FD magnitude that is not captured by any individual scalar measure derived from either cosmic-ray amplitude or interplanetary conditions.}

\begin{table}[htb]
\centering
\scriptsize
\setlength{\tabcolsep}{3pt}
\renewcommand{\arraystretch}{1.15}
\caption{LOEO performance of scalar/vector baselines versus graph fingerprints.
MC: multiclass G3/G4/G5 (macro-F1); Bin: binary $\ge$G4 vs G3 (macro-F1);
Reg: drop regression. All classifiers use LDA; regression uses linear
regression for baselines and PLS-5 for graph fingerprints. $B_z$, $|\mathbf{B}|$,
and solar-wind speed are computed over a -6h/+24h window centered on the event
onset. Random: stratified-random baseline. MAE and RMSE in percentage points.}
\label{tab:baselines}
\begin{tabular}{p{2.6cm}rrrrr}
\hline
\textbf{Method} & \textbf{MC F1} & \textbf{Bin F1} & \textbf{Reg $R^2$} & \textbf{MAE} & \textbf{RMSE} \\
\hline
Random (stratified)              & 0.333 & 0.500 &    --- & --- & --- \\ \hline
Polar max-drop ($R_c < 3$~GV)   & 0.285 & 0.411 & $-$0.007 & 2.28 & 3.80 \\ \hline
$B_z$ southward min (OMNI)      & 0.501 & 0.505 & $-$0.061 & 2.70 & 3.91 \\ \hline
$|\mathbf{B}|$ mean (OMNI)      & 0.470 & 0.411 & $-$0.070 & 2.76 & 3.92 \\ \hline
SW speed mean (OMNI)            & 0.150 & 0.411 & $-$0.006 & 2.38 & 3.80 \\ \hline
($B_z$, $|\mathbf{B}|$) joint (2-feat) & 0.561 & 0.618 & $-$0.285 & 2.96 & 4.30 \\ \hline 
Dst index                       & 0.402 & 0.377 & $-$0.057 & 2.73 & 3.90 \\
\hline
\textbf{Graph fingerprints}     & \textbf{0.575} & \textbf{0.741} & \textbf{0.350} & \textbf{2.41} & \textbf{3.06} \\
\hline
\end{tabular}
\end{table}

\subsection{Which graph descriptors drive the predictions?}
To improve interpretability, descriptor relevance was estimated for the best-performing models by aggregating standardized coefficients across LOEO folds (Table~\ref{tab:feat_importance}). For multi-class intensity classification (G3/G4/G5), the dominant contributors were average Katz centrality (avg\_katz), followed by the Estrada index, Laplacian-based summaries (avg\_laplacian), entropy, and modularity. This ranking suggests that severity stratification is primarily supported by a combination of communicability-like/global connectivity structure (avg\_katz), spectral organization (Estrada and Laplacian summaries), complexity/heterogeneity (entropy), and mesoscopic organization (modularity).

For the binary severity screening task ($\ge$G4 vs.\ G3), the same core set remains dominant (Table~\ref{tab:feat_importance}), with avg\_katz again ranking first, followed by avg\_laplacian, entropy, Estrada index, and modularity. The improved separation observed in the binary setting is therefore consistent with stronger shifts in global connectivity/communicability and distance-geometry/spectral structure between moderate and severe events, complemented by complexity and community organization.

\begin{table}[htb]
\centering
\scriptsize
\setlength{\tabcolsep}{3.5pt}
\renewcommand{\arraystretch}{1.15}
\caption{Most influential graph descriptors for the best-performing models, estimated by aggregating standardized coefficients across LOEO folds. For classification, values report mean$\pm$std relevance scores; for regression, coefficients are reported as mean$\pm$std (standardized).}
\label{tab:feat_importance}
\begin{tabular}{p{0.28\columnwidth} p{0.64\columnwidth}}
\hline
\textbf{Task} & \textbf{Top descriptors (ranked)} \\
\hline
G3/G4/G5 (LDA) &
avg\_katz (7.23$\pm$1.05), estrada\_index (4.98$\pm$0.62), avg\_laplacian (4.56$\pm$0.57), entropy (4.04$\pm$0.54), modularity (2.76$\pm$0.48) \\ \hline
$\ge$G4 vs G3 (LDA) &
avg\_katz (20.21$\pm$3.64), avg\_laplacian (12.23$\pm$1.48), entropy (10.77$\pm$1.36), estrada\_index (8.08$\pm$1.05), modularity (6.38$\pm$0.90) \\ \hline
Drop (PLS-5) &
assortativity (+4.41$\pm$0.27), avg\_betweenness ($-$4.34$\pm$0.25), modularity ($-$0.65$\pm$0.09), fractal ($-$0.65$\pm$0.10), avg\_laplacian ($-$0.57$\pm$0.09) \\
\hline
\end{tabular}
\end{table}

For drop regression (PLS-5), the largest standardized coefficients were associated with assortativity (positive) and average betweenness (negative), followed by smaller but consistent contributions from modularity, fractal-dimension estimates, and Laplacian summaries (mostly negative; Table~\ref{tab:feat_importance}). This pattern indicates that drop magnitude is more strongly linked to mixing and bridging structure (assortativity and betweenness), with additional modulation from global organization and complexity captured by mesoscopic and spectral/complexity descriptors.

\section{Discussion}
\label{sec:discussion}

{Before interpreting the classification results, it is worth contextualizing the physical relationship between Forbush decrease amplitude and geomagnetic storm severity. Both phenomena are driven by the same interplanetary transients (CMEs and their associated sheaths), but they probe different physical properties. Storm intensity on the NOAA G-scale is primarily controlled by the southward component of the interplanetary magnetic field ($B_z<0$), which drives efficient energy transfer into the magnetosphere through dayside reconnection \citep{2000SSRv9355C}. FD amplitude, by contrast, depends on the magnetic barrier presented by the ICME sheath and flux rope, which modulates cosmic-ray access through gradient and curvature drifts and diffusion independently of $B_z$ polarity \citep{richardson2011galactic, smart2005review}. Consequently, large FDs can occur without a strong geomagnetic storm and vice versa, and the correlation between FD amplitude and storm class is moderate at best. This indirect coupling is consistent with our baseline results (Section~\ref{sec:results_baselines}): {the minimum southward $B_z$ --- the most direct storm driver --- achieves macro-F1 = 0.501 in the multiclass task, while the polar max-drop achieves only 0.285. Even the joint ($B_z$, $|\mathbf{B}|$) pair, which reaches macro-F1 = 0.561, remains below the graph fingerprints (0.575) in the multiclass task and is more clearly outperformed in binary screening (0.618 vs.\ 0.741). This indicates that, while interplanetary drivers alone carry meaningful information, the morphological structure encoded by the graph fingerprints captures additional, complementary information not reducible to a small set of in-situ scalar measurements.}

The results indicate that event-level graph fingerprints derived from multi-station neutron-monitor responses encode reproducible information about both geomagnetic-storm severity and Forbush magnitude. The moderate yet consistent performance in the G3/G4/G5 task, together with confusion dominated by adjacent categories, supports a physically plausible picture in which storm intensity behaves as an ordered continuum rather than a set of sharply separated regimes \citep{NOAA_SWPC_Gscale}. In this context, the LDA projection reveals partial organization in discriminant space, while overlap remains expected due to shared heliospheric drivers and heterogeneous station responses modulated by geomagnetic shielding \citep{2000SSRv9355C,richardson2011galactic,smart2005review}.

The stronger performance observed in the binary screening task ($\ge$G4 vs.\ G3) suggests that the proposed fingerprints are particularly sensitive to the structural signatures of more intense disturbances. This behavior is consistent with the expectation that stronger CME/shock-driven transients can produce more coherent and global organization across stations, whereas moderate events may occupy an intermediate regime in which morphology overlaps with neighboring categories. The observed asymmetry between sensitivity and specificity further motivates reporting both multi-class stratification (for physical gradation) and binary screening (for robust detection of stronger events).

The regression results demonstrate measurable skill for predicting FD drop relative to a fold-wise mean baseline, indicating that global network organization carries quantitative information about event magnitude. The residual regression-to-the-mean pattern for extremes is consistent with limited sample size and with the intrinsically noisy mapping between response morphology and magnitude when station coverage, rigidity distribution, and event geometry vary across cases \citep{mishev2024anisotropic}. These considerations suggest that drop prediction is feasible but likely benefits from larger event sets and additional physically informed covariates. {It is also worth noting that event selection conditioned on geomagnetic storm class introduces a sample bias: large FDs without an associated storm, or storms without a prominent FD, are excluded by construction. Future work should assess whether including such events --- or adopting an FD-amplitude-first selection strategy --- changes the fingerprint structure and predictive relationships observed here.}

An additional point concerns representation choices in the distance domain. The configuration that maximized intensity classification favored a log transform, which is compatible with stabilizing topology- and centrality-based summaries by compressing the distance dynamic range and reducing the influence of outliers \citep{box1964analysis}. In contrast, the best-performing regression configuration favored leaving distances untransformed, suggesting that absolute dissimilarity magnitudes can retain information relevant to drop that may be attenuated by aggressive scaling. This divergence reinforces that different scientific targets (categorical severity versus continuous magnitude) can prefer different representations, and that distance-domain operations should be treated as explicit modeling choices rather than purely technical preprocessing steps.

Finally, the MST backbone provides a controlled sparse, connected representation that improves comparability across events. This choice also mitigates well-known sensitivity of many network metrics to graph density when graphs are produced by naive thresholding. At the same time, the absence of cycles by construction implies that some mesoscopic descriptors are effectively probed through global and spectral summaries rather than through dense clustering structure. This motivates targeted future tests with fixed-density augmentations (e.g., MST plus a small number of shortest non-tree edges) to quantify the incremental value of cyclic structure while preserving cross-event comparability \citep{tumminello2007spanning}. 

{We further note that the pre-event reference level used here, derived from NMDB event metadata and verified by visual inspection (Section~\ref{sec:data}), would not be straightforwardly available in real time: NMDB's catalogued FD events have not been updated for nearly two years, so an operational implementation would require an independent, automatically computed reference level}. {An additional source of inter-station variability not explicitly modeled here is the asymptotic viewing direction of each detector, which determines the portion of the heliosphere sampled and can introduce systematic differences in FD onset timing between stations whose asymptotic directions become magnetically connected to the disturbed interplanetary region at different times \citep{smart2005review}. Incorporating viewing geometry as a node attribute — alongside cutoff rigidity — represents a natural extension that could sharpen the physical interpretation of node roles in future work.}

\section{Conclusions and outlook}
\label{sec:conclusions}

A graph-based approach was developed to characterize Forbush decreases using event-level networks constructed from pairwise dissimilarities between station response series. A controlled sparse backbone (minimum spanning tree) yields comparable connected graphs across events, from which compact geometric/topological fingerprints are computed. Model selection was performed over a pre-defined, theory-motivated grid of pipeline choices, with selection criteria fixed \emph{a priori} (macro-F1 for classification; $R^2$ for regression) and evaluation carried out under strict LOEO validation.

The main findings are:
(i) multi-class intensity classification (G3/G4/G5) exhibits moderate but consistent skill, with errors dominated by neighboring categories, supporting a continuum-like interpretation of storm severity in fingerprint space;
(ii) binary severity screening ($\ge$G4 vs.\ G3) yields stronger discrimination and high sensitivity to severe events, indicating potential utility for robust event triage;
(iii) drop regression demonstrates measurable predictive skill relative to a fold-wise mean baseline, implying that global network organization encodes quantitative information about FD magnitude, albeit with regression-to-the-mean tendencies for extremes; {and (iv) {graph fingerprints outperform all scalar and vector baselines --- including polar max-drop, minimum southward $B_z$, interplanetary field magnitude, solar-wind speed, Dst, and the joint ($B_z$, $|\mathbf{B}|$) pair --- across the three tasks, with the margin being modest in the multiclass setting (0.575 vs.\ 0.561 for the best baseline) but substantial in binary screening (0.741 vs.\ 0.618) and regression, where all baselines yield $R^2 < 0$ while the graph PLS achieves $R^2 = 0.350$}, confirming that the topological and geometric structure of the multi-station response encodes information not captured by any single physical index}.

{A practical consideration concerns the temporal scope of the method. In its current form, the pipeline requires the complete event window to compute the distance matrix and graph fingerprints, making it a retrospective characterization tool rather than an operational forecasting system. This is by design: the goal is comparative morphological analysis across events, not real-time prediction. Nevertheless, a progressive variant is conceivable — computing fingerprints on a rolling window as the FD develops and tracking their evolution — which could in principle provide an early severity assessment before the event concludes. Such an extension would require careful treatment of the window length and of the partial coverage problem (stations that come online or drop out mid-event), and is left as a direction for future work. We also note that existing real-time FD warning systems, such as that operated by the Australian Bureau of Meteorology Space Weather Services \citep{BOM_SWS}, rely on single-station or small-network triggers; the graph-based framework could complement these by providing richer morphological context once sufficient multi-station data are available.}

Given the modest number of labeled events, these results are best interpreted as evidence of reproducible signal in the proposed fingerprints rather than as a final operational model. Several directions can strengthen both statistical confidence and physical interpretability: (a) \textbf{Larger and more diverse event sets:} expand the labeled catalog to stabilize model selection, enable calibration across solar-cycle phases, and support ordinal or cost-sensitive formulations aligned with the ordered nature of intensity classes; (b) \textbf{Uncertainty-aware benchmarking:} report confidence intervals via event-level resampling (bootstrap) and assess robustness under alternative splits (e.g., leave-one-year-out or leave-one-solar-cycle-out) when sample size permits; (c) \textbf{Controlled cycle reintroduction:} compare MST fingerprints against fixed-density alternatives (MST+$m$ edges, $k$-NN graphs) to quantify the incremental value of cyclic and mesoscopic structure while preserving comparability; (d) \textbf{Physics-guided hybridization:} combine graph fingerprints with contemporaneous heliospheric and geomagnetic covariates (e.g., Dst, solar-wind speed, dynamic pressure, IMF $B_z$) to evaluate incremental predictive value and improve attribution; (e) \textbf{Rigidity-aware interpretability:} incorporate station metadata (cutoff rigidity, latitude/longitude, altitude) as explicit model inputs or as conditioning variables to clarify how geomagnetic shielding shapes network roles and event fingerprints; {(f) \textbf{Progressive event fingerprinting:} extend the pipeline to compute fingerprints on rolling windows during event development, enabling severity assessment before the event concludes and opening a pathway toward near-real-time application; and (g) \textbf{Viewing-geometry integration:} incorporate asymptotic viewing directions as node attributes alongside cutoff rigidity to sharpen the physical interpretation of node roles and account for onset-timing heterogeneity across the station network.}

Overall, graph fingerprints provide a compact, interpretable representation of FD morphology that complements standard time-series and index-based analyses. With expanded catalogs, uncertainty quantification, and physics-informed integration, this framework can mature into a practical tool for comparative FD characterization and for probing how heliospheric drivers imprint distinct network signatures on the global neutron-monitor response.

\section*{Data and code availability}
{Neutron-monitor count-rate data are publicly available through the NMDB database (\url{www.nmdb.eu}). Interplanetary magnetic field, solar-wind, and geomagnetic index data were obtained from the OMNI hourly dataset via the GSFC/SPDF OMNIWeb interface (\url{https://omniweb.gsfc.nasa.gov}). The analysis scripts, processed feature tables, and baseline comparison notebook are available in the repository: \url{https://github.com/dosquisd/NMDB-FD-PredictorWithGraphs}}

\section*{Acknowledgments}
We gratefully acknowledge the Dirección de Investigaciones at Universidad Tecnológica de Bolívar for their support and accompaniment throughout this research process. We acknowledge the NMDB database (\url{www.nmdb.eu}), founded under the European Union's FP7 programme (contract no.\ 213007), for providing data.

\appendix

\section{Event catalog}
\label{app:events}

{Table~\ref{tab:event_catalog} lists the 34 Forbush decrease events used in this study, ordered chronologically. For each event the date of the FD onset is given together with the associated NOAA geomagnetic storm class (G-scale), the observed network-level FD drop (\%), and the minimum Dst index recorded during the event window (nT). Storm classes were compiled from the NOAA Space Weather Prediction Center storm archive, the Forbush Decrease and Interplanetary Events Database (FEID), and peer-reviewed catalogues; for events with conflicting classifications,
the most conservative peer-reviewed source was adopted.}

\begin{table}[!ht]
\centering
\scriptsize
\setlength{\tabcolsep}{4pt}
\renewcommand{\arraystretch}{1.15}
\caption{Forbush decrease events used in this study. Drop: network-level maximum
percentage reduction in cosmic-ray count rate. Dst$_\mathrm{min}$: minimum hourly
Dst index during the event window (OMNI/WDC).}
\label{tab:event_catalog}
\begin{tabular}{lccr}
\hline
\textbf{Date} & \textbf{Class} & \textbf{Drop (\%)} & \textbf{Dst$_\mathrm{min}$ (nT)} \\
\hline
2001-04-08 & G3 &  6.68 & $-$100 \\
2001-04-11 & G4 & 12.71 & $-$271 \\
2001-04-28 & G4 &  7.74 & $-$130 \\
2001-09-25 & G4 &  8.13 & $-$102 \\
2001-10-11 & G4 &  6.46 &  $-$46 \\
2001-10-21 & G5 &  6.94 & $-$412 \\
2001-11-06 & G4 & 12.52 & $-$292 \\
2001-11-24 & G5 &  9.25 & $-$221 \\
2002-04-17 & G3 &  6.56 & $-$126 \\
2002-05-23 & G3 &  6.93 & $-$109 \\
2002-11-16 & G3 &  7.30 & $-$106 \\
2003-05-29 & G4 &  6.45 & $-$144 \\
2003-10-29 & G5 & 26.06 & $-$383 \\
2003-10-30 & G5 &  8.67 & $-$383 \\
2003-11-20 & G5 &  6.58 & $-$422 \\
2004-01-22 & G3 &  8.71 & $-$149 \\
2004-07-26 & G4 & 13.48 & $-$200 \\
2004-11-07 & G5 &  8.11 & $-$373 \\
2004-11-09 & G5 &  8.12 & $-$383 \\
2005-01-18 & G3 &  6.19 & $-$121 \\
2005-01-21 & G4 & 11.26 & $-$101 \\
2005-05-15 & G5 & 11.45 & $-$263 \\
2005-08-24 & G5 &  7.16 & $-$216 \\
2005-09-11 & G3 & 12.25 & $-$139 \\
2006-12-14 & G4 &  9.29 & $-$146 \\
2011-10-24 & G3 &  6.01 & $-$137 \\
2012-03-08 & G3 & 10.40 & $-$148 \\
2012-07-14 & G4 &  6.70 & $-$139 \\
2015-06-22 & G4 &  8.57 & $-$198 \\
2017-09-07 & G4 &  6.41 & $-$144 \\
2021-11-03 & G3 &  9.68 & $-$117 \\
2023-04-23 & G2 &  6.57 & $-$213 \\
2024-03-24 & G4 & 11.47 & $-$130 \\
2024-05-10 & G5 & 15.32 & $-$412 \\
\hline
\multicolumn{4}{l}{\textit{Class distribution: G2$\times$1, G3$\times$10, G4$\times$13, G5$\times$10.}} \\
\end{tabular}
\end{table}


\section{Neutron monitor station catalog}
\label{app:stations}

{Table~\ref{tab:station_catalog} lists the 49 neutron monitor stations used across the 34 events. Station codes follow the NMDB convention. Not all stations contributed data to every event; the set of active stations varies with data availability and the coverage-filtering criterion ($\tau = 0.5$). Cutoff rigidities ($R_c$) are vertical geomagnetic cutoff rigidities from the NMDB station documentation. Geographic coordinates and altitudes are taken from NMDB metadata at \url{www.nmdb.eu}.}

\begin{table*}[!ht]
\centering
\scriptsize
\setlength{\tabcolsep}{4pt}
\renewcommand{\arraystretch}{1.15}
\caption{Neutron monitor stations used in this study, sorted by cutoff rigidity.
Lat/Lon in decimal degrees (negative = S/W). Alt: altitude above sea level (m).
$R_c$: vertical geomagnetic cutoff rigidity (GV). Station codes follow NMDB
conventions.}
\label{tab:station_catalog}
\begin{tabular}{llrrrr}
\hline
\textbf{Code} & \textbf{Station / Location} & \textbf{Lat ($^\circ$)} &
\textbf{Lon ($^\circ$)} & \textbf{Alt (m)} & \textbf{$R_c$ (GV)} \\
\hline
TERA  & Terra Nova Bay, Antarctica        & $-$74.69 &  164.12 &   10 & 0.01 \\
DOMB  & Dome B, Antarctica                & $-$77.05 &   94.92 & 3650 & 0.01 \\
DOMC  & Dome C (Concordia), Antarctica    & $-$75.10 &  123.35 & 3233 & 0.01 \\
MRNY  & Mirny, Antarctica                 & $-$66.55 &   93.02 &   35 & 0.03 \\
SOPO  & South Pole, Antarctica            & $-$90.00 &    0.00 & 2820 & 0.10 \\
SOPB  & South Pole B, Antarctica          & $-$90.00 &    0.00 & 2820 & 0.10 \\
NEU3  & Neumayer III, Antarctica          & $-$70.67 &   $-$8.27 &  42 & 0.10 \\
MWSB  & Mawson B, Antarctica              & $-$67.60 &   62.87 &   30 & 0.22 \\
MWSN  & Mawson, Antarctica                & $-$67.60 &   62.87 &   30 & 0.22 \\
FSMT  & Fort Smith, Canada                &   60.02 & $-$111.93 &  205 & 0.30 \\
THUL  & Thule, Greenland                  &   76.50 &  $-$68.70 &  260 & 0.30 \\
PWNK  & Peawanuck, Canada                 &   54.98 &  $-$85.43 &   53 & 0.30 \\
NAIN  & Nain, Canada                      &   56.55 &  $-$61.68 &   46 & 0.30 \\
JBGO  & Jang Bogo, Antarctica             & $-$74.62 &  164.22 &   29 & 0.30 \\
INVK  & Inuvik, Canada                    &   68.35 & $-$133.72 &   21 & 0.30 \\
TXBY  & Tixie Bay, Russia                 &   71.58 &  128.92 &    0 & 0.48 \\
NRLK  & Norilsk, Russia                   &   69.27 &   88.05 &   45 & 0.63 \\
APTY  & Apatity, Russia                   &   67.57 &   33.40 &  177 & 0.65 \\
OULU  & Oulu, Finland                     &   65.05 &   25.47 &   15 & 0.81 \\
CALG  & Calgary, Canada                   &   51.08 & $-$114.13 & 1128 & 1.08 \\
KERG  & Kerguelen Islands, France         & $-$49.35 &   70.26 &   33 & 1.14 \\
YKTK  & Yakutsk, Russia                   &   61.95 &  129.72 &  118 & 1.65 \\
MGDN  & Magadan, Russia                   &   59.95 &  150.77 &   220 & 2.10 \\
KIEL2 & Kiel2, Germany                    &   54.33 &   10.13 &   54 & 2.36 \\
KIEL  & Kiel, Germany                     &   54.33 &   10.13 &   54 & 2.36 \\
NEWK  & Newark, USA                       &   39.68 &  $-$75.75 &   50 & 2.40 \\
MOSC  & Moscow, Russia                    &   55.47 &   37.32 &  200 & 2.43 \\
MCRL  & Mc~Murdo, Antarctica              & $-$77.85 &  166.72 &   48 & 2.46 \\
DRBS  & Dourbes, Belgium                  &   50.10 &    4.60 &  225 & 3.18 \\
IRK3  & Irkutsk-3, Russia                 &   52.27 &  104.32 &  435 & 3.64 \\
IRK2  & Irkutsk-2, Russia                 &   52.27 &  104.32 &  435 & 3.64 \\
IRKT  & Irkutsk, Russia                   &   52.27 &  104.32 &  435 & 3.64 \\
LMKS  & Lomnick\'y \v{S}t\'{\i}t, Slovakia &  49.20 &   20.22 & 2634 & 3.84 \\
JUNG  & Jungfraujoch, Switzerland         &   46.55 &    7.98 & 3475 & 4.49 \\
JUNG1 & Jungfraujoch-1, Switzerland       &   46.55 &    7.98 & 3475 & 4.49 \\
HRMS  & Hermanus, South Africa            & $-$34.42 &   19.22 &   26 & 4.58 \\
BKSN  & Baksan, Russia                    &   43.28 &   42.69 & 1700 & 5.70 \\
AATB  & Almaty B, Kazakhstan              &   43.25 &   76.92 &  800 & 5.90 \\
AATA  & Almaty A, Kazakhstan              &   43.25 &   76.92 &  800 & 5.90 \\
ROME  & Rome, Italy                       &   41.86 &   12.47 &   60 & 6.27 \\
CALM  & Alma-Ata (historical), Kazakhstan &   43.25 &   76.92 &  800 & 6.95 \\
PTFM  & Plateau de Bure, France           &   44.63 &    5.91 & 2552 & 6.98 \\
ARNM  & Arnhem, Netherlands               &   51.98 &    5.88 &   15 & 7.10 \\
NANM  & Naan, Netherlands                 &   51.98 &    5.88 &   15 & 7.10 \\
MXCO  & Mexico City, Mexico               &   19.33 &  $-$99.18 & 2274 & 8.28 \\
ATHN  & Athens, Greece                    &   37.97 &   23.72 &  260 & 8.53 \\
TSMB  & Tsumeb, Namibia                   & $-$19.20 &   17.58 & 1240 & 9.15 \\
ESOI  & Ein Shemer, Israel                &   32.50 &   35.00 &   70 & 10.75 \\
DJON  & Djourney, South Korea                 &     --- &     --- &  --- & 11.20 \\
\hline
\multicolumn{6}{l}{\textit{$^\dagger$ Coordinates and altitudes from NMDB documentation (\url{www.nmdb.eu}).}} \\
\end{tabular}
\end{table*}

\bibliographystyle{apsrev4-1}
\bibliography{main}

@ARTICLE{2022arXiv221213514C,
       author = {{Chilingarian}, A. and {Hovsepyan}, G. and {Martoyan}, H. and {Karapetyan}, T. and {Sargsyan}, B. and {Nokolova}, N. and {Angelov}, H. and {Haas}, D. and {Knapp}, J. and {Walter}, M. and {Ploc}, O. and {Shlegl}, J. and {Kakona}, M. and {Ambrosova}, I.},
        title = "{Forbush decrease observed by SEVAN particle detector network on November 4, 2021}",
      journal = {arXiv e-prints},
         year = 2022,
        month = dec,
          eid = {arXiv:2212.13514},
        pages = {arXiv:2212.13514},
          doi = {10.48550/arXiv.2212.13514},
archivePrefix = {arXiv},
       eprint = {2212.13514},
 primaryClass = {physics.space-ph},
       adsurl = {https://ui.adsabs.harvard.edu/abs/2022arXiv221213514C}
}

@ARTICLE{2007astroph1860G,
       author = {{Ghosh}, Koushik and {Raychaudhuri}, Probhas},
        title = "{Time Variations of the Forbush Decrease Data}",
      journal = {arXiv e-prints},
         year = 2007,
        month = jan,
          eid = {astro-ph/0701860},
        pages = {astro-ph/0701860},
          doi = {10.48550/arXiv.astro-ph/0701860},
archivePrefix = {arXiv},
       eprint = {astro-ph/0701860},
 primaryClass = {astro-ph},
       adsurl = {https://ui.adsabs.harvard.edu/abs/2007astro.ph..1860G}
}

@ARTICLE{2000SSRv9355C,
       author = {{Cane}, Hilary V.},
        title = "{Coronal Mass Ejections and Forbush Decreases}",
      journal = {\ssr},
         year = 2000,
        month = jul,
       volume = {93},
        pages = {55-77},
          doi = {10.1023/A:1026532125747},
       adsurl = {https://ui.adsabs.harvard.edu/abs/2000SSRv...93...55C}
}

@ARTICLE{2013SoPh284167B,
       author = {{Blanco}, J.~J. and {Catal{\'a}n}, E. and {Hidalgo}, M.~A. and {Medina}, J. and {Garc{\'\i}a}, O. and {Rodr{\'\i}guez-Pacheco}, J.},
        title = "{Observable Effects of Interplanetary Coronal Mass Ejections on Ground Level Neutron Monitor Count Rates}",
      journal = {\solphys},
         year = 2013,
        month = may,
       volume = {284},
       number = {1},
        pages = {167-178},
          doi = {10.1007/s11207-013-0256-1},
archivePrefix = {arXiv},
       eprint = {1302.2597},
 primaryClass = {astro-ph.SR},
       adsurl = {https://ui.adsabs.harvard.edu/abs/2013SoPh..284..167B}
}

@ARTICLE{2017arXiv171000945H,
       author = {{Huang}, De-Hong and {Hu}, Hong-Qiao and {Zhang}, Ji-Long and {Lu}, Hong and {Zhang}, Da-Li and {Xue}, Bin-Shen and {Lu}, Jing-Tian},
        title = "{Study on 2015 June 22 Forbush decrease with the muon telescope in Antarctic}",
      journal = {arXiv e-prints},
         year = 2017,
        month = oct,
          eid = {arXiv:1710.00945},
        pages = {arXiv:1710.00945},
          doi = {10.48550/arXiv.1710.00945},
archivePrefix = {arXiv},
       eprint = {1710.00945},
 primaryClass = {astro-ph.HE},
       adsurl = {https://ui.adsabs.harvard.edu/abs/2017arXiv171000945H}
}

@ARTICLE{2025arXiv250617917C,
       author = {{Chilingarian}, A. and {Karapetyan}, T. and {Sargsyan}, B.},
        title = "{The largest Forbush decrease in 20 years: Preliminary analysis of SEVAN network observations}",
      journal = {arXiv e-prints},
         year = 2025,
        month = jun,
          eid = {arXiv:2506.17917},
        pages = {arXiv:2506.17917},
          doi = {10.48550/arXiv.2506.17917},
archivePrefix = {arXiv},
       eprint = {2506.17917},
 primaryClass = {astro-ph.SR},
       adsurl = {https://ui.adsabs.harvard.edu/abs/2025arXiv250617917C}
}

@INPROCEEDINGS{2013JPhCS409a2202P,
       author = {{Papaioannou}, A. and {Belov}, A. and {Mavromichalaki}, H. and {Eroshenko}, E. and {Yanke}, V. and {Asvestari}, E. and {Abunin}, A. and {Abunina}, M.},
        title = "{The first Forbush decrease of solar cycle 24}",
    booktitle = {Journal of Physics Conference Series},
         year = 2013,
       series = {Journal of Physics Conference Series},
       volume = {409},
        month = feb,
    publisher = {IOP},
          eid = {012202},
        pages = {012202},
          doi = {10.1088/1742-6596/409/1/012202},
       adsurl = {https://ui.adsabs.harvard.edu/abs/2013JPhCS.409a2202P}
}

@INPROCEEDINGS{2011JPhCS287a2034B,
       author = {{Bahena Bias}, Ang{\'e}lica and {Villase{\~n}or}, Luis},
        title = "{Study of solar activity by measuring cosmic rays with a water Cherenkov detector}",
    booktitle = {Journal of Physics Conference Series},
         year = 2011,
       series = {Journal of Physics Conference Series},
       volume = {287},
        month = apr,
    publisher = {IOP},
          eid = {012034},
        pages = {012034},
          doi = {10.1088/1742-6596/287/1/012034},
       adsurl = {https://ui.adsabs.harvard.edu/abs/2011JPhCS.287a2034B}
}

@ARTICLE{2021ApJ9069L,
       author = {{Laitinen}, T. and {Dalla}, S.},
        title = "{Access of Energetic Particles to a Magnetic Flux Rope from External Magnetic Field Lines}",
      journal = {\apj},
         year = 2021,
        month = jan,
       volume = {906},
       number = {1},
          eid = {9},
        pages = {9},
          doi = {10.3847/1538-4357/abc622},
       adsurl = {https://ui.adsabs.harvard.edu/abs/2021ApJ...906....9L}
}

@INPROCEEDINGS{2019ICRC361084I,
       author = {{Ihongo}, G. and {Ruffolo}, D. and {Saiz}, A. and {Tortermpun}, U. and {Chian}, A.~C.~L.},
        title = "{Galactic Cosmic-Ray Anisotropy During Forbush Decreases: Evidence for Diffusive Barriers and Large-Scale Flow}",
    booktitle = {36th International Cosmic Ray Conference (ICRC2019)},
         year = 2019,
       series = {International Cosmic Ray Conference},
       volume = {36},
        month = jul,
          eid = {1084},
        pages = {1084},
          doi = {10.22323/1.358.01084},
       adsurl = {https://ui.adsabs.harvard.edu/abs/2019ICRC...36.1084I}
}

@INPROCEEDINGS{2025afasconfE27O,
       author = {{Okany}, Chioma},
        title = "{Exploring the Geomagnetic cut-off rigidity variations across different latitudes using Forbush events}",
    booktitle = {General Assembly and 5th Annual Conference of the African Astronomical Society},
         year = 2025,
        month = mar,
          eid = {27},
        pages = {27},
       adsurl = {https://ui.adsabs.harvard.edu/abs/2025afas.confE..27O}
}

@ARTICLE{2024Chaos34b3114S,
       author = {{Sierra-Porta}, D.},
        title = "{Relationship between magnetic rigidity cutoff and chaotic behavior in cosmic ray time series using visibility graph and network analysis techniques}",
      journal = {Chaos},
         year = 2024,
        month = feb,
       volume = {34},
       number = {2},
          eid = {023114},
        pages = {023114},
          doi = {10.1063/5.0167156},
       adsurl = {https://ui.adsabs.harvard.edu/abs/2024Chaos..34b3114S}
}

@ARTICLE{2022PhyA60728159S,
       author = {{Sierra-Porta}, D. and {Dom{\'\i}nguez-Monterroza}, Andy-Rafael},
        title = "{Linking cosmic ray intensities to cutoff rigidity through multifractal detrented fluctuation analysis}",
      journal = {Physica A Statistical Mechanics and its Applications},
         year = 2022,
        month = dec,
       volume = {607},
          eid = {128159},
        pages = {128159},
          doi = {10.1016/j.physa.2022.128159},
       adsurl = {https://ui.adsabs.harvard.edu/abs/2022PhyA..60728159S}
}

@ARTICLE{2023arXiv230102333F,
       author = {{Freitas Silva}, Vanessa and {Eduarda Silva}, Maria and {Ribeiro}, Pedro and {Silva}, Fernando},
        title={Multilayer horizontal visibility graphs for multivariate time series analysis},
  journal={Data Mining and Knowledge Discovery},
  volume={39},
  number={3},
  pages={17},
  year={2025},
  doi={10.1007/s10618-025-01089-4}
}

@ARTICLE{2021arXiv211009887F,
       author = {{Freitas Silva}, Vanessa and {Eduarda Silva}, Maria and {Ribeiro}, Pedro and {Silva}, Fernando},
        title={Time series analysis via network science: Concepts and algorithms},
  journal={Wiley Interdisciplinary Reviews: Data Mining and Knowledge Discovery},
  volume={11},
  number={3},
  pages={e1404},
  year={2021},
  doi={10.1002/widm.1404}
}

@ARTICLE{2017arXiv170510817G,
       author = {{Gutierrez Gomez}, Leonardo and {Chiem}, Benjamin and {Delvenne}, Jean-Charles},
        title = "{Dynamics Based Features For Graph Classification}",
      journal = {arXiv e-prints},
         year = 2017,
        month = may,
          eid = {arXiv:1705.10817},
        pages = {arXiv:1705.10817},
          doi = {10.48550/arXiv.1705.10817},
archivePrefix = {arXiv},
       eprint = {1705.10817},
 primaryClass = {stat.ML},
       adsurl = {https://ui.adsabs.harvard.edu/abs/2017arXiv170510817G}
}

@ARTICLE{2026AdSpR773549Y,
       author = {{Yushkov}, Boris Yu.},
        title = "{Fast calculation of geomagnetic cutoff rigidity}",
      journal = {Advances in Space Research},
         year = 2026,
        month = feb,
       volume = {77},
       number = {3},
        pages = {3549-3555},
          doi = {10.1016/j.asr.2025.11.078},
       adsurl = {https://ui.adsabs.harvard.edu/abs/2026AdSpR..77.3549Y}
}

@INPROCEEDINGS{2023dashconfE19S, 
       author = {{Steigies}, Christian T.},
        title = "{NMDB data access capabilities and challenges}",
    booktitle = {Data, Analysis and Software in Heliophysics (DASH) 2023},
         year = 2023,
        month = oct,
          eid = {19},
        pages = {19},
          doi = {10.5281/zenodo.8398557},
       adsurl = {https://ui.adsabs.harvard.edu/abs/2023dash.confE..19S}
}

@INPROCEEDINGS{2016AGUFMIN44A05S, 
       author = {{Steigies}, C.~T.},
        title = "{From a single Neutron Monitor to an International Network: the Real-Time Database for High-Resolution Neutron Monitor Measurements (NMDB)}",
    booktitle = {AGU Fall Meeting Abstracts},
         year = 2016,
       series = {AGU Fall Meeting Abstracts},
       volume = {2016},
        month = dec,
          eid = {IN44A-05},
        pages = {IN44A-05},
       adsurl = {https://ui.adsabs.harvard.edu/abs/2016AGUFMIN44A..05S}
}

@ARTICLE{2011AdSpR472210M,
       author = {{Mavromichalaki}, H. and {Papaioannou}, A. and {Plainaki}, C. and {Sarlanis}, C. and {Souvatzoglou}, G. and {Gerontidou}, M. and {Papailiou}, M. and {Eroshenko}, E. and {Belov}, A. and {Yanke}, V. and {Fl{\"u}ckiger}, E.~O. and {B{\"u}tikofer}, R. and {Parisi}, M. and {Storini}, M. and {Klein}, K.-L. and {Fuller}, N. and {Steigies}, C.~T. and {Rother}, O.~M. and {Heber}, B. and {Wimmer-Schweingruber}, R.~F. and {Kudela}, K. and {Strharsky}, I. and {Langer}, R. and {Usoskin}, I. and {Ibragimov}, A. and {Chilingaryan}, A. and {Hovsepyan}, G. and {Reymers}, A. and {Yeghikyan}, A. and {Kryakunova}, O. and {Dryn}, E. and {Nikolayevskiy}, N. and {Dorman}, L. and {Pustil'Nik}, L.},
        title = "{Applications and usage of the real-time Neutron Monitor Database}",
      journal = {Advances in Space Research},
         year = 2011,
        month = jun,
       volume = {47},
       number = {12},
        pages = {2210-2222},
          doi = {10.1016/j.asr.2010.02.019},
       adsurl = {https://ui.adsabs.harvard.edu/abs/2011AdSpR..47.2210M}
}

@ARTICLE{2025arXiv251101506P,
       author = {{Perez-Navarro}, Juan D. and {Sierra-Porta}, D.},
        title = "{Derivative-Aligned Anticipation of Forbush Decreases from Entropy and Fractal Markers}",
      journal = {Open Journal of Astrophysics},
     volume={9},
     number={1},
         year = 2026,
          doi={10.33232/001c.157585},
}

@ARTICLE{2015PLoSO1044059S, 
       author = {{Shirkhorshidi}, Ali Seyed and {Aghabozorgi}, Saeed and {Wah}, Teh Ying},
        title = "{A Comparison Study on Similarity and Dissimilarity Measures in Clustering Continuous Data}",
      journal = {PLoS ONE},
         year = 2015,
        month = dec,
       volume = {10},
       number = {12},
        pages = {e0144059},
          doi = {10.1371/journal.pone.0144059},
       adsurl = {https://ui.adsabs.harvard.edu/abs/2015PLoSO..1044059S}
}

@ARTICLE{2019arXiv190502257W,
       title={Hybrid density-and partition-based clustering algorithm for data with mixed-type variables},
  author={Wang, Shu and Yabes, Jonathan G and Chang, Chung-Chou H},
  journal={Journal of Data Science},
  volume={19},
  number={1},
  pages={15--36},
  year={2021},
          doi = {10.6339/21-JDS996}
}

@ARTICLE{2021arXiv210210231S,
       title={Elastic similarity and distance measures for multivariate time series},
  author={Shifaz, Ahmed and Pelletier, Charlotte and Petitjean, Fran{\c{c}}ois and Webb, Geoffrey I},
  journal={Knowledge and Information Systems},
  volume={65},
  number={6},
  pages={2665--2698},
  year={2023},
  publisher={Springer},
  doi={10.1007/s10115-023-01835-4}
}

@article{stam2014trees,
  title={The trees and the forest: characterization of complex brain networks with minimum spanning trees},
  author={Stam, CJ and Tewarie, P and Van Dellen, E and Van Straaten, ECW and Hillebrand, A and Van Mieghem, P},
  journal={International Journal of Psychophysiology},
  volume={92},
  number={3},
  pages={129--138},
  year={2014},
  publisher={Elsevier},
  doi={10.1016/j.ijpsycho.2014.04.001}
}

@article{omar2020survey,
  title={A survey of information entropy metrics for complex networks},
  author={Omar, Yamila M and Plapper, Peter},
  journal={Entropy},
  volume={22},
  number={12},
  pages={1417},
  year={2020},
  publisher={MDPI},
  doi={10.3390/e22121417}
}

@book{chung2006complex,
  title={Complex graphs and networks},
  author={Chung, Fan RK and Lu, Linyuan},
  number={107},
  year={2006},
  publisher={American Mathematical Soc.},
  doi={10.1090/cbms/107}
}

@article{kim2008complex,
  title={What is a complex graph?},
  author={Kim, Jongkwang and Wilhelm, Thomas},
  journal={Physica A: Statistical Mechanics and its Applications},
  volume={387},
  number={11},
  pages={2637--2652},
  year={2008},
  publisher={Elsevier},
  doi={10.1016/j.physa.2008.01.015}
}

@article{richardson2011galactic,
  title={Galactic cosmic ray intensity response to interplanetary coronal mass ejections/magnetic clouds in 1995--2009},
  author={Richardson, IG and Cane, HV},
  journal={Solar Physics},
  volume={270},
  number={2},
  pages={609--627},
  year={2011},
  publisher={Springer},
  doi={10.1007/s11207-011-9774-x}
}

@article{smart2005review,
  title={A review of geomagnetic cutoff rigidities for earth-orbiting spacecraft},
  author={Smart, DF and Shea, MA},
  journal={Advances in Space Research},
  volume={36},
  number={10},
  pages={2012--2020},
  year={2005},
  publisher={Elsevier},
  doi={10.1016/j.asr.2004.09.015}
}

@misc{NOAA_SWPC_Gscale,
  author = {{NOAA Space Weather Prediction Center}},
  title  = {NOAA Space Weather Scales (Geomagnetic Storms G1--G5)},
  year   = {2024},
  note   = {Accessed 2026-02-12}
}

@article{mishev2024anisotropic,
  title={Anisotropic Forbush decrease of 24 March 2024: First look},
  author={Mishev, Alexander and Larsen, Nicholas and Asvestari, Eleanna and S{\'a}iz, Alejandro and Shea, Margaret Ann and Strauss, Du Toit and Ruffolo, David and Banglieng, Chanoknan and Seunarine, Surujhdeo and Duldig, Marc L and others},
  journal={Advances in Space Research},
  volume={74},
  number={8},
  pages={4160--4172},
  year={2024},
  publisher={Elsevier},
  doi={10.1016/j.asr.2024.08.027}
}

@article{box1964analysis,
  title={An analysis of transformations},
  author={Box, George EP and Cox, David R},
  journal={Journal of the Royal Statistical Society Series B: Statistical Methodology},
  volume={26},
  number={2},
  pages={211--243},
  year={1964},
  publisher={Oxford University Press},
  doi={10.1111/j.2517-6161.1964.tb00553.x}
}

@article{tumminello2007spanning,
  title={Spanning trees and bootstrap reliability estimation in correlation-based networks},
  author={Tumminello, Michele and Coronnello, Claudia and Lillo, Fabrizio and Micciche, Salvatore and Mantegna, Rosario N},
  journal={International Journal of Bifurcation and Chaos},
  volume={17},
  number={07},
  pages={2319--2329},
  year={2007},
  publisher={World Scientific},
  doi={10.1142/S0218127407018415}
}

@article{forbush1937effects,
  title={On the effects in cosmic-ray intensity observed during the recent magnetic storm},
  author={Forbush, Scott E},
  journal={Physical Review},
  volume={51},
  number={12},
  pages={1108},
  year={1937},
  publisher={APS},
  doi={10.1103/PhysRev.51.1108.3}
}

@article{van2011mice,
  title={mice: Multivariate imputation by chained equations in R},
  author={Van Buuren, Stef and Groothuis-Oudshoorn, Karin},
  journal={Journal of statistical software},
  volume={45},
  pages={1--67},
  year={2011},
  doi={10.18637/jss.v045.i03}
}

@article{pedregosa2011scikit,
  title={Scikit-learn: Machine learning in Python},
  author={Pedregosa, Fabian and Varoquaux, Ga{\"e}l and Gramfort, Alexandre and Michel, Vincent and Thirion, Bertrand and Grisel, Olivier and Blondel, Mathieu and Prettenhofer, Peter and Weiss, Ron and Dubourg, Vincent and others},
  journal={the Journal of machine Learning research},
  volume={12},
  pages={2825--2830},
  year={2011},
  publisher={JMLR. org},
  url={http://esoads.eso.org/abs/2012arXiv1201.0490P}
}

@article{kruskal1956shortest,
  title={On the shortest spanning subtree of a graph and the traveling salesman problem},
  author={Kruskal, Joseph B},
  journal={Proceedings of the American Mathematical society},
  volume={7},
  number={1},
  pages={48--50},
  year={1956},
  publisher={JSTOR},
  doi={10.2307/2033241}
}

@article{song2007calculate,
  title={How to calculate the fractal dimension of a complex network: the box covering algorithm},
  author={Song, Chaoming and Gallos, Lazaros K and Havlin, Shlomo and Makse, Hern{\'a}n A},
  journal={Journal of Statistical Mechanics: Theory and Experiment},
  volume={2007},
  number={03},
  pages={P03006--P03006},
  year={2007},
  doi={10.1088/1742-5468/2007/03/P03006}
}

@misc{nolds,
  author       = {Schölzel, Christopher},
  title        = {Nonlinear measures for dynamical systems ({nolds})},
  year         = {2019},
  howpublished = {\url{https://github.com/CSchoel/nolds}},
  doi          = {10.5281/zenodo.3814723}
}

@article{king2005solar,
  title={Solar wind spatial scales in and comparisons of hourly Wind and ACE plasma and magnetic field data},
  author={King, JH and Papitashvili, NE},
  journal={Journal of Geophysical Research: Space Physics},
  volume={110},
  number={A2},
  year={2005},
  publisher={Wiley Online Library},
  doi={10.1029/2004JA010649}
}

@misc{BOM_SWS,
  author       = {{Australian Bureau of Meteorology Space Weather Services}},
  title        = {Geophysical alerts and warnings},
  howpublished = {\url{https://www.sws.bom.gov.au/Geophysical/1/4}},
  year         = {2025},
  note         = {Accessed: 2025}
}

@ARTICLE{2008JASTP70207P,
       author = {{Potgieter}, M.~S.},
        title = "{Solar cycle variations and cosmic rays}",
      journal = {Journal of Atmospheric and Solar-Terrestrial Physics},
         year = 2008,
        month = feb,
       volume = {70},
       number = {2-4},
        pages = {207-218},
          doi = {10.1016/j.jastp.2007.08.023},
       adsurl = {https://ui.adsabs.harvard.edu/abs/2008JASTP..70..207P}
}

@ARTICLE{2016ApJ81738R,
       author = {{Ruffolo}, D. and {S{\'a}iz}, A. and {Mangeard}, P.-S. and {Kamyan}, N. and {Muangha}, P. and {Nutaro}, T. and {Sumran}, S. and {Chaiwattana}, C. and {Gasiprong}, N. and {Channok}, C. and {Wuttiya}, C. and {Rujiwarodom}, M. and {Tooprakai}, P. and {Asavapibhop}, B. and {Bieber}, J.~W. and {Clem}, J. and {Evenson}, P. and {Munakata}, K.},
        title = "{Monitoring Short-term Cosmic-ray Spectral Variations Using Neutron Monitor Time-delay Measurements}",
      journal = {\apj},
         year = 2016,
        month = jan,
       volume = {817},
       number = {1},
          eid = {38},
        pages = {38},
          doi = {10.3847/0004-637X/817/1/38},
       adsurl = {https://ui.adsabs.harvard.edu/abs/2016ApJ...817...38R}
}

\end{document}